\documentclass{ecai} 
\usepackage{latexsym}
\usepackage{amssymb}
\usepackage{amsthm}
\usepackage{booktabs}
\usepackage{enumitem}
\usepackage{makecell} % Required package

\usepackage[T1]{fontenc}
\usepackage[utf8]{inputenc}
\usepackage{graphicx}
\usepackage{xspace}
\usepackage{color}
\usepackage{enumitem}
\usepackage[linesnumbered,ruled,vlined]{algorithm2e}
\usepackage{amsmath}
\usepackage{multirow}

\newcommand{\methodName}{RuleGenie\xspace}

\newcommand{\BibTeX}{B\kern-.05em{\sc i\kern-.025em b}\kern-.08em\TeX}

\begin{document}

\begin{frontmatter}

%%% Use this command to specify your submission number.
%%% In doubleblind mode, it will be printed on the first page.

\paperid{9987} 

\title{\methodName: SIEM Detection Rule Set Optimization}

\author{Akansha Shukla, Parth Atulbhai Gandhi, Yuval Elovici and Asaf Shabtai}

\address{Beg-Gurion University of the Negev, Israel}

\begin{abstract}
Security information and event management (SIEM) systems serve as a critical hub, employing rule-based logic to detect and respond to threats. 
Redundant or overlapping rules in SIEM systems lead to excessive false alerts, degrading analyst performance due to alert fatigue, and increase computational overhead and response latency for actual threats.
As a result, optimizing SIEM rule sets is essential for efficient operations.
Despite the importance of such optimization, research in this area is limited, with current practices relying on manual optimization methods that are both time-consuming and error-prone due to the scale and complexity of enterprise-level rule sets.
To address this gap, we present \methodName, a novel large language model (LLM) aided recommender system designed to optimize SIEM rule sets. 
Our approach leverages transformer models' multi-head attention capabilities to generate SIEM rule embeddings, which are then analyzed using a similarity matching algorithm to identify the top-k most similar rules. 
The LLM then processes the rules identified, utilizing its information extraction, language understanding, and reasoning capabilities to analyze rule similarity, evaluate threat coverage and performance metrics, and deliver optimized recommendations for refining the rule set. 
By automating the rule optimization process, \methodName allows security teams to focus on more strategic tasks while enhancing the efficiency of SIEM systems and strengthening organizations' security posture. 
We evaluated \methodName on a comprehensive set of real-world SIEM rule formats, including Splunk, Sigma, and AQL (Ariel query language), demonstrating its platform-agnostic capabilities and adaptability across diverse security infrastructures. 
Our experimental results show that \methodName can effectively identify redundant rules, which in turn decreases false positive rates and enhances overall rule efficiency.
\end{abstract}
\end{frontmatter}
%\keywords{SIEM, Rule optimization, Large language model.}

\section{Introduction\label{sec:intro}}

In today’s rapidly evolving digital landscape, enterprises must implement robust security measures to combat increasingly sophisticated cyber threats~\cite{hiscox2024cyber}. 
As organizations scale, their IT infrastructure correspondingly expands, encompassing a range of components such as networking devices, cloud computing environments, and endpoints. 
This expansion of IT infrastructure generates a diverse and vast stream of security events.

To cope with the increasing volume of security data, enterprises depend heavily on security information and event management (SIEM) systems~\cite{SIEM}. 
SIEM systems are specifically designed to handle the challenges of processing diverse, high-volume data by aggregating events from multiple sources, each of which may use a unique vendor-specific schema. The ability of SIEM systems to standardize diverse data formats into a unified representation plays a vital role in modern enterprise security operations, enabling streamlined event analysis, storage, and incident response~\cite{SIEM_rep}.

The operational core of SIEM architecture is anchored in its sophisticated rule-based system, where specialized rule engines serve as a critical bridge between raw security data and actionable intelligence~\cite{SIEM_rules}. 
This rule-based processing capability enables security operations center (SOC) analysts to detect and respond to threats more effectively.
Therefore, the efficacy of a SIEM's detection and response capabilities depends on well-structured rule sets that eliminate redundancy, maximize coverage, and ensure maintainability.

Organizations continuously integrate new rules into their SIEM environment to adapt to emerging threats, accommodate new infrastructure components, and comply with regulatory requirements. 
While adopting an incremental approach to threat detection can enhance coverage, it also introduces significant operational challenges. 
As the amount of detection rules increases, overlapping logic, redundant conditions, and conflicting alerts may emerge, placing a burden on SOC analysts, who must filter through redundant alerts, a process consuming critical time and resources. 
The resulting alert fatigue can compromise analysts' ability to recognize genuine threats~\cite{fatigue}. 
This increases the risk that serious security incidents will be undetected, threatening the organization's overall security posture.

Addressing theses challenges requires a systematic approach to optimizing rule sets and eliminating redundancies. 
Currently, SIEM system rule optimization relies heavily on manual efforts by analysts who must sift through entire rule sets to identify redundancies, merge similar rules, and remove unnecessary rules. 

This process, which is both time-consuming and prone to human error, requires significant expertise and experience, resulting in the inefficient allocation of human resources and an increased workload for analysts. 
This ultimately affects the overall efficiency of cybersecurity operations. 
While both SQL and SIEM systems fundamentally deal with pattern matching and data filtering at scale, but the SIEM's operational contexts differ significantly. 
The extensive research performed on SQL query optimization underscores the significance of automated maintenance for optimized rule sets. 

Traditional databases prioritize static, structured data, whereas SIEM systems manage real-time, heterogeneous security data streams, introducing complexities that extend beyond the scope of conventional SQL optimization methods. 
This unique technical distinction underscores the need for SIEM rule optimization, yet despite the critical operational importance of optimized SIEM rule sets in cybersecurity infrastructure, research in this domain remains notably limited.

To bridge this research gap, we present \methodName, a novel human-in-the-loop (HITL) system~\cite{nunes2015survey} that performs both retrospective and prospective analysis of SIEM rules. 
The proposed system retrospectively analyzes existing rule sets for potential redundancies while prospectively recommending optimizations when new rules are introduced, thereby continuously ensuring rule set efficiency.

\methodName implements a robust pipeline to optimize and remove duplicate SIEM rules. 
It leverages an encoder-decoder transformer model~\cite{egonmwan} to generate embeddings of rules, enabling efficient syntactic representation. 
\methodName employs a distance-based similarity detection algorithm~\cite{qian2004similarity} to identify potentially redundant or overlapping rules in relation to a target rule in a rule set. 
The target rule may represent a newly proposed rule or an existing rule from the rule set being analyzed. 

Each rule pair, consisting of the target rule and a potentially similar rule identified through similarity matching is then analyzed using a structured chain-of-thought (CoT)~\cite{wei2022chain} reasoning process performed by an LLM. 
This CoT analysis is performed in three key steps. 
First, the LLM evaluates functional and semantic overlap between the rules. 
Second, it assesses platform specificity and hierarchical relationships between target environments. 
Finally, it compares the rules based on operational metrics including coverage, the false positive rate, and computational efficiency. 
By integrating these aspects of analysis, \methodName produces data-driven recommendations for rule set optimization, ensuring that no blind spots emerge in an organization's threat detection capabilities while maintaining SIEM systems' operational efficiency.

\methodName was assessed across diverse SIEM rule formats, including Sigma~\cite{sigma2023}, Splunk~\cite{splunkDocs}, and AQL (Ariel query language) rule definitions. 
The evaluation encompassed 2,347 Sigma rules and 1,640 Splunk security content rules, with respectively 139 and 58 redundant rules demonstrating overlapping functionalities expressed via  identical query structures or detection objectives. 
To validate the platform-agnostic nature of our approach, we converted the 2,347 Sigma rules into AQL security rules used by QRadar and examined \methodName's ability to identify redundant rule pairs across multiple SIEM environments. 
In addition, we evaluated \methodName using two LLMs: a proprietary model and a local open-source model. 
The proprietary model achieved over 90\% recall and 65\% precision across all SIEM rule formats.
The local open-source model achieved over 80\% recall and 75\% precision across all SIEM rule formats, while also ensuring data privacy and reduced costs.

\noindent The main contributions of this paper can be summarized as follows:
\begin{itemize}[itemsep=0.1em, topsep=0pt]
    \item \textbf{A novel application of LLMs for SIEM rule set optimization:}  
    We introduce a context-aware LLM-based approach for detecting similar rules and providing recommendations that reduces redundancies while maintaining high precision in evolving SIEM environments. 
    \item \textbf{A code-based embedding technique for syntactic comparison of SIEM rules:} We propose the use of advanced code-based embedding techniques to enable syntactic comparisons of SIEM rules, ensuring more accurate and context-rich rule matching.
    \item \textbf{Real-world applicability:} We rigorously tested and validated \methodName in real-world SIEM environments, demonstrating its effectiveness in handling large-scale rule sets typical of enterprise deployments. 
\end{itemize}

\section{Related Work\label{sec:related}}

To our knowledge, no prior work has directly addressed SIEM rule optimization, even though rule redundancy has long been recognized as a critical issue in cybersecurity.

Existing research in related areas such as firewall rule management, where research has shown that rule redundancy minimization significantly reduce noise and improve system efficiency~\cite{liu2008complete,liu2008all,persson2020optimization}. 
Unlike firewall rules, which are relatively simple, SIEM rules involve intricate event aggregation, data transformation, and multi-source data correlation logic. 
Consequently, the optimization approaches that work well for firewall rules are often unsuitable for SIEM environments. 
But these studies reveal key insight that more security rules does not necessarily correlate with better security, but can instead introduce performance degradation and increase the likelihood of false positives.
These methods, while effective for well-defined rule sets, face significant limitations when applied to the complex logic of SIEM rules.

Recent work has demonstrated the applicability of LLMs in a range of cybersecurity tasks, including log analysis~\cite{saha2024llm, lee, han2023loggpt, liu2024, guo2021logbert, shield}, and modeling threat behavior~\cite{garza2023assessing}. 
LLMs have shown the ability to understand and manipulate structured queries and configurations, including security rule formats like Sigma. However, their potential for automated rule optimization in operational security contexts particularly for identifying redundancy, semantic overlap, or logical inefficiencies in detection rules remains largely unexplored.

\section{Methodology\label{sec:method}}

\subsection{\methodName Overview}

\begin{figure*}[h]
    \makebox[\textwidth][c]{%
        \includegraphics[width=0.95\linewidth]{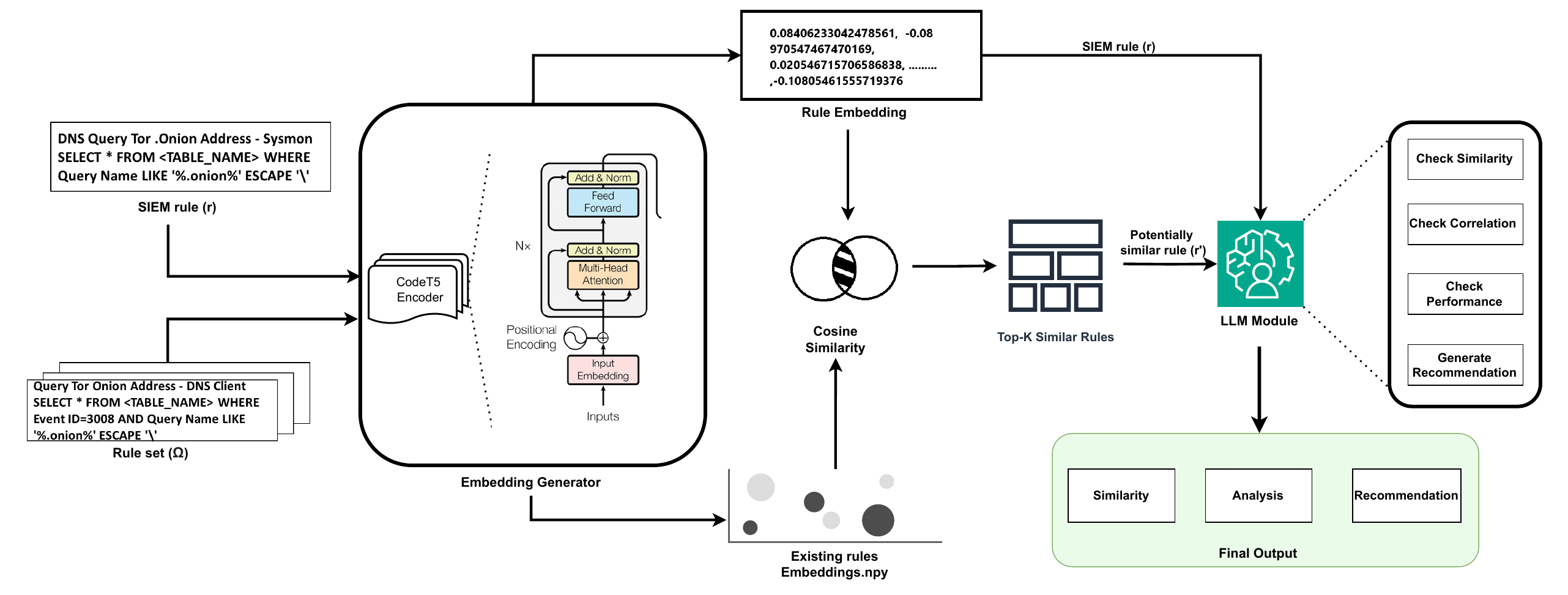}%
    }
    \caption{An overview of \methodName's three-phase pipeline.}
    \label{fig:pipeline}
\end{figure*}

The \methodName pipeline comprises three phases: rule embedding generation, similarity detection, and LLM analysis, as illustrated in Figure~\ref{fig:pipeline}. 
In the first phase, SIEM rules are transformed into embedding representations in a high-dimensional vector to facilitate syntactical analysis. 
In the second phase, the vectorized embedding representations of target rule and existing rule set is compared using similarity matching algorithm. 
Once a target rule has been compared to an existing rule set, a cluster of potentially similar rules is generated. 
In the final phase, LLM-aided analysis is performed to evaluate both the target rule and the cluster of similar rules. 

The \methodName pipeline leverages transformer models for SIEM rule vectorization and uses LLMs for semantic analysis. 

\subsection{Rule Embedding Generation\label{subsec:rule_emb}}

In the initial phase, \methodName utilizes a transformer-based architecture to generate rule embeddings. This embedding process forms the cornerstone of the pipeline, ensuring that downstream tasks benefit from comprehensive and context-rich representations of SIEM rules and enhancing the overall pipeline's effectiveness. In the  subsections below, we describe the model selection and validation framework, implementation architecture, and overflow management pipeline, all of which are employed when generating transformer embeddings.

\subsubsection{Model Selection and Validation}

We conducted a comparative analysis of state-of-the-art transformer models to identify the most suitable architecture for encoding SIEM rule syntax. 
The evaluation included general-purpose language models (BERT~\cite{bert}), security-specific model (SecBERT~\cite{secbert}), and code-oriented models (CodeT5~\cite{codet5}).

The vector distance analysis between redundant Sigma rules demonstrated that code-based models achieve superior performance in maintaining vector proximity for syntactically similar rules.
CodeT5 emerged as the optimal choice for our pipeline based on several key attributes:

\begin{enumerate}
    \item \textbf{Architecture design:} Its pretrained encoder-decoder transformer architecture, derived from T5, excels at both comprehension and generation tasks.
    
    \item \textbf{Token processing:} Its identifier-aware training objective enables precise token distinction and recovery.
    
    \item \textbf{Dual-modality training:} Its training on both code and natural language facilitates effective alignment between natural and programming language constructs.
\end{enumerate}

\subsubsection{Implementation Architecture}

Our embedding framework employs an adaptive segmentation strategy coupled with a conditional processing pipeline to accommodate SIEM rules of varying complexity while ensuring consistent vector representations. 
The architecture implements two processing mechanisms based on input dimensionality:

\begin{itemize}
    \item \textit{Standard processing mechanism.}
The \methodName pipeline uses an advanced segmentation approach to handle SIEM rules that exceed standard token length limits (512-4096 tokens). 
The system employs a sophisticated chunking mechanism designed to preserve logical boundaries within each segment. 
After segmentation, each partition is processed independently through the encoder, which captures the salient features of the input. These encoded segments are then recombined to preserve the relationships between chunks and maintain the overall context of the rule.

    \item \textit{Overflow management mechanism.}
To maintain models' robustness when processing exceptionally complex SIEM rules (exceeding 4096 tokens), we employ a graceful degradation ~\cite{herlihy1991specifying} protocol that initiates a controlled fallback to a near null-space representation.
In critical security contexts, these instances are flagged for optional manual review.This overflow management mechanism ensures that system integrity is upheld even under extreme load conditions.
\end{itemize}

This dual-architecture for embedding generation ensures consistent processing capabilities across SIEM rules of varying  complexity while preserving the dimensional structure of the vector space representation. 
The system design prioritizes both processing efficiency and representation uniformity, making it well-suited for security operations.

\subsection{Similarity Detection\label{subsec:similarity}}

In this phase of the pipeline, cosine similarity~\cite{li2013distance} is used to quantify the syntactic relationships between SIEM rule embeddings. 
 
Cosine similarity was chosen as the metric, because it is agnostic to vector magnitude, making it particularly robust for comparing embeddings of varying complexities.

To operationalize this similarity measure, we implement a top-$k$ retrieval mechanism (in our evaluations we set $k$ = 5) for each SIEM rule. 
The selection of $k$ and its impact on precision and recall is systematically analyzed in Section~\ref{sec:eval}, where we present empirical evidence supporting this choice through an ablation study.

Additionally, our experimental investigation of transformer selection reveals that the embedding space exhibits interesting topological properties, where syntactical related rules form well-defined clusters.
\subsection{Redundancy Analysis Using LLM}\label{subsec:llm}
%\vspace{-1.6em}
The last phase of our framework leverages an LLM enhanced with CoT reasoning to perform fine-grained semantic and functional analysis of security rules. 
We initially used GPT-4o ~\cite{gpt}, which demonstrated exceptional performance in understanding complex security rule semantics and generating nuanced recommendations. 
However, operational constraints—particularly data privacy requirements and cost considerations necessitated the exploration of alternative solutions.
Our subsequent evaluation of open-source models focused on three criteria: (1) context length capacity for handling extensive rule descriptions; (2) analytical capabilities for semantic analysis; and (3) computational efficiency and resource optimization.

After testing multiple candidates, including Llama 8B fine-tuned \newline~\cite{llama} and Qwen models ~\cite{qwen}, we selected Qwen-2.5-14B-Instruct as our primary LLM. 
This model demonstrated superior performance in our specific security rule analysis tasks, offering an optimal balance between computational efficiency and analytical capability.

To maximize the effectiveness of the Qwen-2.5-14B-Instruct model, we developed prompting strategies that leverage its context understanding capabilities and a standardized JSON response format that structures the model's outputs across all analytical stages while maintaining computational efficiency. 
This structured approach ensures consistent, machine-readable outputs while preserving the sophisticated analytical capabilities required for security rule optimization.

The LLM-driven redundancy analysis consists of four stages which are described below.

\subsubsection{Deep semantic analysis and functional mapping}
The process begins with the LLM evaluating a pair of rules $(r, r')$, where $r$ represents the current rule and $r'$ represents a candidate from the $k$-similar neighbors identified using the similarity matching algorithm. 
The LLM performs semantic analysis, considering both syntactic structure and functional intent. The rules are compared by LLM for structural matching, semantic equivalence and conditional overlap  necessary to establish whether $r$ and $r'$  exhibit a significant degree of similarity. This analysis yields two key outputs: (a) a boolean classification indicating semantic and functional overlap; and (b) similarity score $s \in [0, 100]$, where 0 indicates no similarity and 100 indicates perfect similarity.

Through evaluation on Sigma rules, we established a minimum similarity threshold of 75, which effectively identifies significant rule overlaps while minimizing false positives as discussed in Section \ref{sec:sim_thres}. Rules with $s \geq 75$ proceed to subsequent analysis stages, while others are pruned to maintain computational efficiency and ensure meaningful comparisons.
\subsubsection{Hierarchical dependence recognition}
For qualifying rule pairs, we employ a novel hierarchical relationship detection algorithm that examines both structural and semantic dependencies. The analysis focuses on identifying three primary relationship patterns:
\begin{enumerate}
    \item \textbf{Platform-specific independence}, where rules serve similar security objectives but operate within distinct technological contexts.
    \item \textbf{Generalization relationships}, where one rule subsumes or extends another functionality. For example, consider the Splunk rules Abnormally High AWS Instances Terminated by User (AWS-specific) and Abnormally High Cloud Instances Terminated by User (covers AWS, Azure, GCP, etc.). Here, the second rule generalizes the first by applying the same detection logic across multiple cloud providers.
    \item \textbf{Cross-platform dependencies}, which reveal potential optimization opportunities across different security platforms.
\end{enumerate}

This hierarchical mapping enables sophisticated rule organization strategies and forms the foundation for optimization decisions.

\subsubsection{Performance and quality optimization}
In this stage a comprehensive evaluation framework for dependent rule pairs is implemented to analyze their operational effectiveness. 
Comparative evaluation is conducted across three performance metrics:
\begin{enumerate}
    \item \textbf{Coverage Analysis}: The scope and comprehensiveness of each rule's detection capabilities are assessed. Broader coverage is prioritized.
    \item \textbf{Efficiency Metrics}: Execution times and resource consumption are evaluated to minimize operational overhead.
    \item \textbf{False Positive Mitigation}: The potential for false positive generation is rigorously assessed based on three key aspects: conditional statement scope, filtering criteria, and validation logic implementation.  Each rule is further analyzed for its effectiveness in minimizing blind spots and ensuring robust validation logic. After evaluating both the target rule $r$ and candidate rule $r'$ against these criteria, the results are compared to identify the superior rule, with a preference for rules that maintain high accuracy while reducing noise.
\end{enumerate}

\subsubsection{Rule recommendation for SIEM environments}

Integrating insights from semantic analysis, hierarchical dependency mapping, and performance evaluation, the final stage of \methodName focuses on generating finely tuned recommendations. By carefully balancing trade-offs between conflicting metrics such as efficiency and accuracy, \methodName provides systematic and targeted strategies for optimizing SIEM rules.

For each rule pair analyzed, recommendations are generated based on their relationship: \begin{enumerate}
    \item \textbf{Keep superior rule:} If one rule demonstrates superior coverage and false positive handling, the recommendation is to retain the better-performing rule.
    
    \item \textbf{Merge complementary rules:} If a rule pair is similar but not entirely redundant and could offer enhanced coverage when combined, \methodName recommends merging the rules.
    
    \item \textbf{Keep both rules:} For rule pairs that are clearly distinct and serve different detection purposes, the recommendation is to retain both rules.
\end{enumerate}

This structured framework enables precise reasoning that improves rule relevance and operational effectiveness, addressing the challenges of modern cybersecurity environments and resulting in a high-confidence rule deployment strategy designed to be robust, adaptable, and contextually aligned with operational requirements. This ensures that recommendations are not only balanced and actionable but also tailored to the dynamic demands of security operations, fostering advanced threat detection and mitigation.

\section{Evaluation\label{sec:evaluation}}
\subsection{Experimental Setup}

\methodName was implemented using Python 3.11. We preprocess SIEM rules to extract key features. 
Embeddings for these rules are generated using the encoder component of Salesforce's CodeT5 model. 
We employed two large language models for rule similarity comparison, performance analysis, and recommendation generation: OpenAI's GPT-4o, accessed via the Azure OpenAI API, and a local deployment of Qwen-2.5-14B-Instruct, managed using the Hugging Face library. The locally deployed Qwen model can run efficiently on a single NVIDIA RTX 3080 GPU.\footnote{Code will be made available once the paper is accepted.}

\subsection{Dataset}

To address security considerations surrounding sensitive SIEM infrastructure, we limited our experimental analysis to detection rules sourced from public security repositories.
Our data acquisition framework leveraged two prominent open-source security initiatives: the Sigma detection framework~\cite{sigma2023} and Splunk Security Content repository~\cite{splunkDocs}.
The integration of these complementary sources yielded 3,987 detection rules, which we methodically pre-classified into three distinct categories: (1) foundational detection rules for baseline security monitoring, (2) advance threat hunting rules, and (3) emerging threat rules targeted at time-critical vulnerabilities and exploits.

For the initial evaluation of \methodName pipeline Splunk and Sigma rules were used.
Through rigorous analysis of 2,347 Sigma rules, 130 Windows-specific rules were identified as new, and the other 2,217 were incorporated in the existing rule set.
The Splunk Security Content repository contributed an additional 100 new rules, complemented by 1,540 existing rules. 
To validate the platform-agnostic nature of our methodology, we converted all 2,347 Sigma rules into AQL queries. We utilized pySigma~\cite{pysigma}, a Python package maintained by the Sigma repository authors, to convert the Sigma rules into both AQL and Splunk query formats. 
 which are methodically pre-classified into three categories: (1) foundational detection rules for baseline security monitoring, (2) advance threat hunting rules, and (3) emerging threat rules targeted at time-critical vulnerabilities.
 
We partitioned the dataset to simulate real-world task of integrating new detections into an existing SIEM environment. We manually curated 230 rules (130 Sigma, 100 Splunk) representing recent additions typically seen in production environments. These form the \textit{new} evaluation set, while the remaining 3,757 detections constitute the \textit{existing} rule set for comparison. To assess platform-agnostic capabilities, all Sigma detections were programmatically converted into both AQL and Splunk query formats using pySigma~\cite{pysigma}, a Python package maintained by the Sigma repository authors.

\begin{table}[h] % Using [htbp] is generally better than [h]
    \centering
    \small % Use small font size
    \caption{Comparison of SIEM rule sets in terms of redundancy and new rules.}
    \label{table:siem_ruleset} % Changed label slightly to avoid duplicate if needed
    \scalebox{0.78}{ % Apply the same scaling factor
    \begin{tabular}{|l|c|c|c|c|} % Added vertical lines | and kept lcccc alignment (5 columns)
        \hline % Use hline instead of booktabs rules
        % Use thead for header wrapping
        \thead{\bfseries SIEM Rule Set} & \thead{\bfseries Non-Redundant\\\bfseries Existing Rules} & \thead{\bfseries Non-Redundant\\\bfseries New Rules} & \thead{\bfseries Redundant\\\bfseries Existing Rules} & \thead{\bfseries Redundant\\\bfseries New Rules} \\
        \hline % hline after header
        Sigma  & 2167 & 41 & 50 & 89 \\
        
        Splunk & 1517 & 65 & 23 & 35 \\
        
        AQL    & 2167 & 41 & 50 & 89 \\
        \hline % hline at the bottom
    \end{tabular}% Add % to avoid spurious space after tabular inside scalebox
    } % End scalebox
\end{table}

This comprehensive classification and conversion process facilitated the development of a robust input dataset analysis and experimentation. 
Table~\ref{table:siem_ruleset} categorizes SIEM rules into redundant and non-redundant groups across Sigma, Splunk, and AQL datasets, covering both newly introduced and existing rules.

\subsection{Ablation Study}\label{sec:eval}

This subsection details the empirical evaluation used to optimize the configuration of \methodName. We present the selection process for key components, including: the embedding generation model, the similarity matching parameter ($k$), the operational similarity threshold, and the large language model for recommendations. We utilize standard precision and recall metrics throughout this evaluation since established SIEM rule optimization benchmarks are not available.

\subsubsection{Embedding Model Selection}

\begin{figure*}[h]
    \centering
    \includegraphics[width=0.96\linewidth]{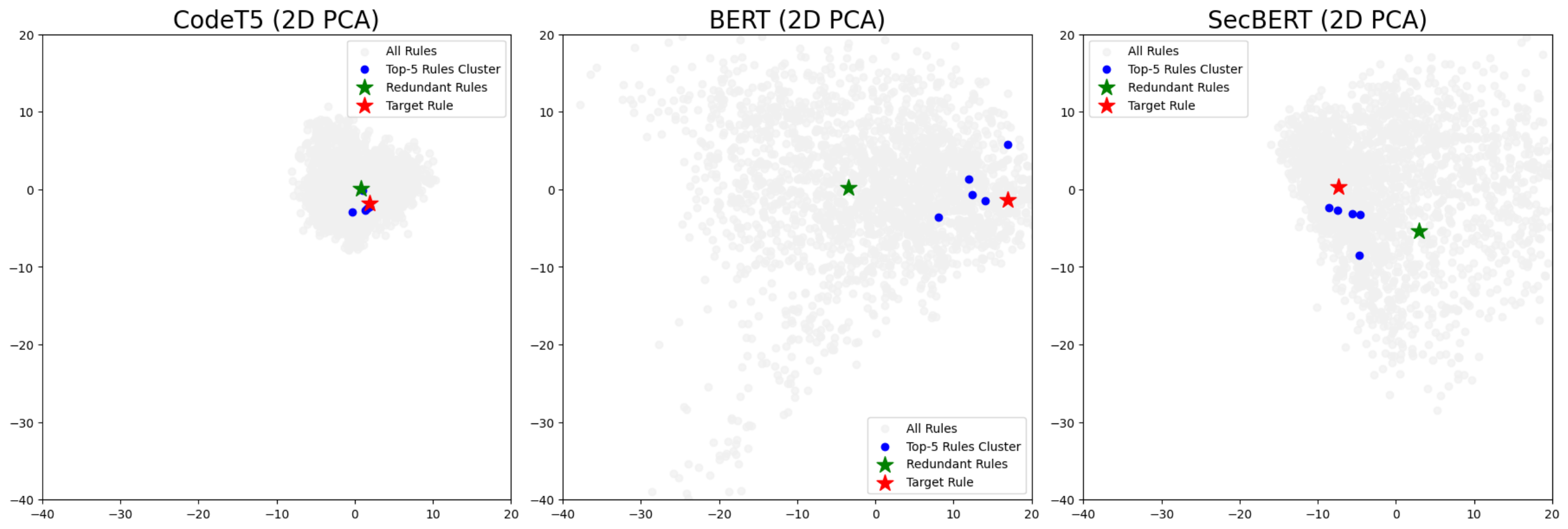}
   \caption{An example of a randomly selected rule. 
   Sigma rule \textit{"New Service Uses Double Ampersand in Path"} embedding using transformer models.}
    \label{fig:transformer}
\end{figure*}

We evaluated three transformer models for generating SIEM rule embeddings: the general-purpose BERT, the security-focused SecBERT, and the code-focused CodeT5. Figure~\ref{fig:transformer} illustrates the performance differences between these models in identifying redundant SIEM rules. The figure displays a 2D representation of the high-dimensional rule embeddings generated by each model, obtained using principal component analysis (PCA).The embedding space projections, has target rule under evaluation marked in red, while the rule that is redundant to target rule is represented in green.The visualizations show CodeT5 embeddings place the redundant rule significantly closer to target rule compared to BERT and SecBERT embeddings. 

This closer proximity indicates that CodeT5 produces embeddings that better capture the similarities relevant for identifying redundancy specifically within SIEM rule syntax and structure. CodeT5's effectiveness likely stems from its pre-training on programming language structures and keywords, which align well with typical SIEM rule syntax, enabling it to represent rule features more effectively for this task than the general-purpose or security-text-focused models.

\subsubsection{Value selection for $k$}

\begin{figure}[h] % Single-column figure
    \centering
    \includegraphics[width=0.9\linewidth]{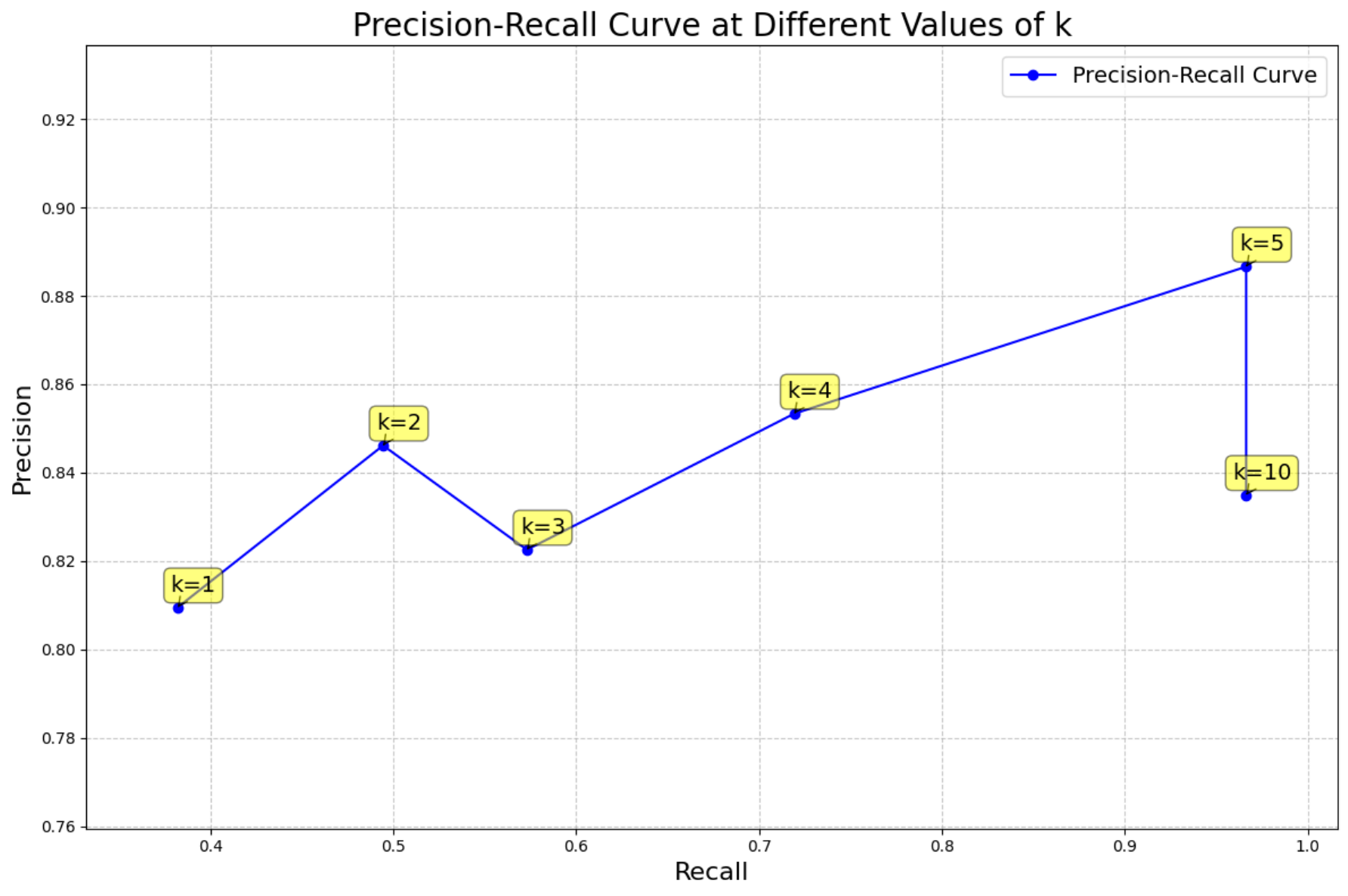}
    \caption{Precision-recall curve for \(k\) in Sigma rule set.}
    \label{fig:kvalue}
\end{figure}

The optimal value of $k$ was empirically determined through analysis of precision-recall trade-offs across different $k$ values for the Sigma rule set as shown in Figure~\ref{fig:kvalue}. 
The results demonstrate that the recall metric values consistently improved as $k$ increases from one to five, with recall values showing substantial improvement from 0.382 to 0.966, representing a 152.94\% increase. 
Similarly, precision showed a generally positive trend up to $k=5$, starting at 0.810 for $k=1$ and reaching its peak of 0.887 at $k=5$, representing a 9.5\% improvement. This simultaneous enhancement in both precision and recall metrics up to $k=5$ indicates strong classifier performance.

While the recall metric values plateau at 0.966 for $k \geq 5$, precision begins to deteriorate for $k > 5$, dropping to 0.843 at $k=10$. 
This represents a 4.9\% decrease in precision with no corresponding gain in recall, suggesting that $k=5$ represents an optimal balance point. The deterioration in precision beyond $k=5$ can be attributed to the inclusion of more distant neighbors in the classification decision, which may introduce noise and reduce the classifier's ability to maintain clear decision boundaries.
\setlength{\parskip}{0pt}
\subsubsection{LLM model selection}

While our analysis confirms that embeddings generated by CodeT5 effectively capture syntatic patterns and locality in SIEM rules with rules sharing similar syntax clustering together (see Figure \ref{fig:transformer}), relying solely on embedding similarity presents several limitations. This syntactic matching lacks a deeper understanding of rule semantics and operational context, risking false positive recommendations or the removal of rules with unique defensive value, potentially creating security blind spots.

To overcome these limitations, we leverage LLMs trained on both natural language and cybersecurity domain knowledge. These models offer a more nuanced interpretation of SIEM rule semantics, considering technical implementation details alongside the security implications of potential modifications. This approach aims to enhance the quality and safety of rule recommendations.

We evaluated three candidate LLMs:

First, we assessed OpenAI's GPT-4o via the Azure API. While demonstrating high accuracy in identifying redundancies and providing contextually relevant recommendations, its operational cost was prohibitive. Processing our complete rule set incurred an estimated cost of \$142.5 per analysis iteration (based on standard API pricing of \$0.03/1K input tokens and \$0.06/1K output tokens). Additionally, the average processing time of 70 minutes made it unsuitable for frequent or large-scale use.

Given the cost constraints of GPT-4o, we next investigated smaller models to assess if a more cost-effective open-source solution could be viable. We experimented with Llama 3.1 8B as a lightweight open-source alternative. However, the base model exhibited a high hallucination rate (70\%) when interpreting rule semantics and operational context. Subsequent fine-tuning yielded only marginal performance improvements. The model’s non linear time complexity proved computationally inefficient for enterprise-scale deployments, particularly when processing large rule sets. This computational constraint, coupled with persistent hallucination issues, rendered the model unsuitable for production environments requiring reliable rule analysis.

Finally, our evaluation focused on Qwen-2.5-14B-Instruct deployed locally. This model demonstrated performance comparable to GPT-4o in identifying relevant rules while satisfying crucial constraints of data privacy via on-premises deployment and operational cost. Specifically, in our rule recommendation task, Qwen achieved a minimum precision of 72\%, and recall of 80\% (see Table~\ref{tab:qwen_siem_ruleset}). These performance metrics, combined with the practical advantages of local deployment, establish Qwen-2.5-14B-Instruct as the most viable LLM solution for \methodName within an enterprise context.

\subsubsection{Similarity Threshold Selection\label{sec:sim_thres}}
To determine the optimal similarity threshold for identifying redundant SIEM rules, we evaluated multiple threshold values using the Sigma rule set, analyzing the resulting precision and recall for each candidate.

As detailed in Table~\ref{tab:sim_performance}, the analysis revealed the expected trade-off: lower thresholds (e.g., 65) increased recall at the cost of precision, while higher thresholds (e.g., 85) improved precision but reduced recall. Based on these results, we selected a threshold of 75 as providing the optimal balance. This value is therefore used as the operational similarity threshold in \methodName.
\setlength{\parskip}{0pt}
\begin{table}[h]
\caption{Performance of \methodName for different similarity score.}
\label{tab:sim_performance}
\centering
\begin{tabular}{|l|l|c|c|}
\hline
\textbf{Similarity Threshold}     & \textbf{Precision} & \textbf{Recall} \\ \hline 
65                    & 0.454             & 0.733          \\ 
75                       & 0.818              & 0.733  \\  
85                       & 0.840              & 0.635          \\ \hline 
\end{tabular}
\end{table}
\setlength{\parskip}{0pt}
\subsection{Evaluation Metrics\label{sec:evaluation_metrics}}

Our evaluation assesses the performance of \methodName pipeline in two stages: (a) the identification of potentially similar or redundant rules, and (b) the quality of the optimization recommendations provided for the identified rules. We evaluate the system’s output against a ground truth established through expert analyst review. 

To quantify \methodName's performance, we utilize metrics that reflect both its ability to accurately detect redundancy and the correctness of its recommendations. While we adopt the terminology of precision and recall for evaluating our method, we define these metrics in a task-specific manner due to the absence of prior baselines.

First, we measure the \methodName's ability to identify truly redundant rules using Recall, consistent with its standard definition in classification tasks. Recall quantifies the proportion of actual redundant rules in the dataset that our system successfully identifies:

\begin{equation}
\resizebox{\columnwidth}{!}{%
  $\displaystyle
    \mathrm{Recall}
    = \frac{\text{\# of correctly identified redundant rules}}
           {\text{\# of correctly identified redundant rules + \# of missed redundant rules}}
  $
}
\end{equation}

Second, to quantify the quality of the optimization recommendations provided by \methodName for the rules it identifies, we define Precision. This metric specifically measures the proportion of recommendations that were validated as correct by human analysts based on three critical matrices: (a) detection coverage; (b) false positive rate reduction; and (c) computational efficiency, out of all the rules the system flagged as potentially redundant:

\begin{equation}
\resizebox{\columnwidth}{!}{%
  $\displaystyle
    \mathrm{Precision}
    = \frac{\text{\# of correct recommendations}}
           {\text{\# of correct recommendations} + \text{\# of incorrect recommendations}}
  $
}
\end{equation}

By quantifying both Precision and Recall, we aim to provide a comprehensive view of the pipeline's effectiveness in supporting human analysts in SIEM rule set optimization.

\section{Results\label{sec:result}}

\subsection{Adding new rules to an existing rule set}

Our experimental evaluation, the results of which are summarized in Table ~\ref{tab:new_rules}, examines the performance of various LLMs in generating security rule recommendations for newly implemented SIEM rules. The results show that Qwen-2.5-14B-Instruct outperforms GPT-4o in terms of precision across all evaluated rule sets, with comparable recall performance between the two models.
\setlength{\parskip}{0pt}
\begin{table}[h]
\caption{Adding new rules to an existing rule set. 
Comparison of models on different rule sets in terms of precision and recall.
Bold numbers indicate best performance and underlined numbers indicate second-best performance.}
\centering
\begin{tabular}{|p{3cm}|l|c|c|}
\hline
\textbf{Model} & \textbf{SIEM Rule Set} & \textbf{Precision} & \textbf{Recall} \\ \hline
GPT-4o  & Sigma   & \underline{0.886}  & \textbf{0.966}   \\ 
    & Splunk      & \underline{0.673}    & \textbf{1.000}    \\ 
     & AQL    & \underline{0.886}     & \textbf{0.966}     \\ \hline
Qwen-2.5-14B-Instruct & Sigma    &  \textbf{0.941}   & \underline{0.910}  \\ 
        & Splunk   & \textbf{0.795}   & \underline{0.886}   \\ 
     & AQL    & \textbf{0.941}    & \underline{0.910}     \\ \hline
Llama 3.1 8B-Instruct Fine-tuned & Sigma & 0.450    & 0.483  \\ 
      & Splunk      & 0.315      & 0.398    \\ 
               & AQL     & 0.450  & 0.483    \\ \hline
\end{tabular}
\label{tab:new_rules}
\end{table}
\setlength{\parskip}{0pt}
\subsection{Optimizing existing rule set}

Based on the cost analysis presented in Section ~\ref{sec:eval}, which revealed significant operational expenses associated with GPT-4o, and the high hallucination rates observed in Llama 3.1 8B, we selected Qwen-2.5-14B-Instruct to optimize the existing SIEM rule set. The results of our evaluation of Qwen-2.5-14B-Instruct's performance in  rule set optimization are presented in Table ~\ref{tab:qwen_siem_ruleset} and demonstrate its effectiveness as a cost-efficient and reliable alternative.
\setlength{\parskip}{0pt}
\begin{table}[h]
\caption{Optimizing existing rule set. Performance of Qwen-2.5-14B-Instruct on different rule sets.}
\label{tab:qwen_siem_ruleset}
\centering
\begin{tabular}{|l|l|c|c|}
\hline
\textbf{Model} & \textbf{SIEM Rule Set} & \textbf{Precision} & \textbf{Recall} \\ \hline
Qwen-2.5-14B-Instruct & Sigma           & 0.947              & 0.850           \\       & Splunk          & 0.726              & 0.804           \\ 
         & AQL             & 0.947              & 0.850           \\ \hline
\end{tabular}
\end{table}
\setlength{\parskip}{0pt}

\section{Discussion\label{sec:discussion}}

\subsection{Impact of Embedding Generation and Similarity Matching}
This study aimed to investigate the effectiveness and efficiency of \methodName for identifying similar rules within a rule set. A key objective is to quantify the contribution of each phase.

To evaluate the impact of our embedding-based pre-processing strategy, we conducted a comparative analysis between two workflows as summarized in Table~\ref{tab:performance}:
\begin{itemize}
  \item \textbf{Baseline Workflow:} Brute-force analysis of SIEM rules, requiring pairwise comparisons across the entire rule set without initial filtering or dimensionality reduction.
  \item \textbf{Proposed Pipeline:} Incorporates the following pre-processing steps: (a) generation of vector embeddings for each rule, (b) retrieval of the top-5 most similar rules via cosine similarity and (c) run downstream analysis exclusively on these top candidate rules
\end{itemize}
\setlength{\parskip}{0pt}

\begin{table}[htbp]
    \centering
    \small
    \caption{Performance comparison between baseline and proposed pipeline (Qwen-2.5-14B-Instruct).}
    \label{tab:performance}
    \scalebox{0.76}{
    \begin{tabular}{|l|c|c|c|c|c|}
        \hline
        \thead{\bfseries Workflow} & \thead{\bfseries Sample\\\bfseries Size} & \thead{\bfseries Total Time\\\bfseries (min)} & \thead{\bfseries Avg. Time\\\bfseries per Rule\\\bfseries (min)} & \thead{\bfseries Throughput\\\bfseries (rules/min)} & \thead{\bfseries Speed‐up\\\bfseries (per-rule)} \\
        \hline
        \makecell[tl]{Baseline \\ (brute force)}
            & 7 & 224 & 32.0 & 0.22 & -- \\
        \hline
        \makecell[tl]{Proposed pipeline \\ (embedding + top‐5)}
            & 100 & 40 & 0.4 & 2.50 & 80$\times$ \\
        \hline
    \end{tabular}%
    }
\end{table}
\setlength{\parskip}{0pt}

The baseline experiment was limited to a sample size of 7 rules due to the prohibitive computational cost of a full 100-rule brute-force analysis. An estimated $100 \times 32.0 = 3200$ min (approximately 53.3h) would have been required, rendering it infeasible within practical time constraints.

The performance improvements shown in Table \ref{tab:performance} derive from embedding-based dimensionality reduction, which compacts the search space, and from similarity pruning, which restricts comparisons to the top five most similar rules thereby mitigating the quadratic time complexity of brute-force, all-pairs evaluation.

\subsection{Enhancing Analysis Quality with Chain-of-Thought}

The second study evaluated the impact of integrating a CoT prompting strategy compared to a naive single-prompt baseline. The results shown in Table \ref{tab:prompt_performance} shows the dramatic increase in evaluation metrics of recall and precision discussed in Section \ref{sec:evaluation_metrics}.

The stepwise reasoning inherent in the CoT methodology enables the model to better understand nuanced relationships between rules, particularly where redundancy stems from semantic equivalence rather than syntactic similarity, thus providing better recommendations. This effectiveness stems from the principle discussed by ~\cite{wei2022chain} that decomposing complex tasks, into simpler steps allows LLMs to process information more reliably and accurately. 

\begin{table}[h]
\caption{Performance of baseline and chain-of-thought prompts (Qwen-2.5-14B-Instruct).}
\label{tab:prompt_performance}
\centering
\begin{tabular}{|l|l|c|c|}
\hline
\textbf{Prompt Type}     & \textbf{Precision} & \textbf{Recall} \\ \hline 
Single prompt                    & 0.250              & 0.533           \\  
CoT prompt                       & 0.818              & 0.733           \\ \hline 
\end{tabular}
\end{table}

In conclusion, the combined contributions of the efficient pre-processing and the sophisticated reasoning are essential components of the \methodName pipeline. The first phase addresses the critical challenge of scalability and computational cost while, second phase addresses the equally important challenge of analytical accuracy and the quality of the recommendations. Together, these phases create a robust and effective system for large-scale rule optimization.

\section{Conclusion and Future Work\label{sec:conclusion}}
In this paper, we introduced \methodName, a novel HITL method that leverages LLMs generated recommendations to optimize SIEM rule sets. Our experimental results demonstrate that \methodName can streamline SOC workflows by minimizing multiple redundant alerts through intelligent rule recommendation.

Building on our findings, future research will focus on two key dimensions: automated implementation of LLM-generated recommendations and optimization of rule execution sequences. This dual approach aims to enhance both the accuracy and computational efficiency of SIEM infrastructure. By developing intelligent orchestration mechanisms for rule execution, we anticipate achieving improved detection capabilities while minimizing system resource utilization. This advancement could significantly enhance an organization's security posture by establishing a more responsive and resource-efficient security monitoring framework.

\bibliography{references}

\end{document}